\documentclass[aps,prl,twocolumn,superscriptaddress,showpacs,a4paper, longbibliography]{revtex4-1}
\usepackage{graphicx}
\usepackage{amssymb,bm}
\usepackage{amsmath}
\usepackage{ulem}
\usepackage{color}
\usepackage{subfigure}
\usepackage{subeqnarray}
\usepackage{wasysym}
\usepackage[pdftex,breaklinks=true,colorlinks=true]{hyperref}
\usepackage{multirow}

\newcommand{\qzero}{$\mathbf q=0$}
\newcommand{\cuboc}{\textit{cuboc1}}

\newcommand{\sqrtsqrt}{$\sqrt3\times\sqrt3$}
\newcommand{\DM}{Dzyaloshinskii-Moriya}
\usepackage[utf8]{inputenc}
\usepackage{comment, currfile, soul}
\usepackage[english]{babel}

\begin{document}

\title{Chiral Spin Liquid on a Kagome Antiferromagnet Induced by \DM\ Interaction}

\author{Laura Messio}
\email{laura.messio@lptmc.jussieu.fr}
\affiliation{Laboratoire de Physique Th\'eorique de la Mati\`ere Condens\'ee, CNRS UMR 7600, Universit\'e Pierre et Marie Curie, Sorbonne Universit\'es, 75252 Paris, France}

\author{Samuel Bieri}
\affiliation{Institute for Theoretical Physics, ETH Z\"urich, 8099 Z\"urich, Switzerland}

\author{Claire Lhuillier}
\affiliation{Laboratoire de Physique Th\'eorique de la Mati\`ere Condens\'ee, CNRS UMR 7600, Universit\'e Pierre et Marie Curie, Sorbonne Universit\'es, 75252 Paris, France}

\author{Bernard Bernu}
\affiliation{Laboratoire de Physique Th\'eorique de la Mati\`ere Condens\'ee, CNRS UMR 7600, Universit\'e Pierre et Marie Curie, Sorbonne Universit\'es, 75252 Paris, France}

\date{\today}

\begin{abstract}
  The quantum spin liquid material herbertsmithite is described by an antiferromagnetic Heisenberg model on the kagome lattice with non-negligible Dzyaloshinskii-Moriya interaction~(DMI).
  A well established phase transition to the $\mathbf q=0$ long-range order occurs in this model when the DMI strength increases, but the precise nature of a small-DMI phase remains controversial.
  Here, we describe a new phase obtained from Schwinger-boson mean-field theory that is stable at small DMI, and which can explain the dispersionless spectrum seen in inelastic neutron scattering experiment by Han {\it et al.}~(\href{http://dx.doi.org/10.1038/nature11659}{Nature (London) {\bf 492}, 406 (2012)}).
  It is a time-reversal symmetry breaking $\mathbb Z_2$ spin liquid, with the unique property of a small and constant spin gap in an extended region of the Brillouin zone.
  The phase diagram as a function of DMI and spin size is given, and dynamical spin structure factors are presented.
\end{abstract}


\maketitle

Frustration in quantum magnets is a captivating and everlasting story.
Competing interactions can lead to unconventional phases such as spin liquids (SL).
After the first proposal by Anderson \cite{Anderson} of a quantum SL in the $S=1/2$ Heisenberg model on the triangular lattice as a zero temperature disordered state, this notion has been greatly refined. A large number of such exotic phases have been discussed, notably on the antiferromagnetic kagome lattice, characterized by fractional symmetry quantum numbers \cite{Qi_Fu, Symmetry_fractionalization}.

Herbertsmithite is a paradigmatic material strongly suspected to host a SL.
It was first synthesized in 2005 \cite{Nocera} and has since been subject to numerous experimental studies \cite{Herbertsmithite_exp_H, Herbert_measurements, Mendels_Kag, DM_Bert, Herbert_measurements2, Nature_DM, Herbert_impurities, Zorko17_PRL.118.017202} (see \cite{Herbert_review} for a recent review).
Herbertsmithite remains disordered down to very low temperatures, and it is described by an antiferromagnetic spin-$1/2$ Heisenberg model on the kagome lattice with strong nearest-neighbor interaction, $H_0 = J\sum_{\langle i,j\rangle} \mathbf S_i\cdot \mathbf S_j$, $J \simeq 200$~K.

In view of the various proposed ground states, it appears that the low-energy physics is quite rich and that even small deformations of this idealized Hamiltonian can have crucial effects. Several perturbations are known to exist. Impurities are physically unavoidable \cite{Herbert_impurities} and theoretically challenging \cite{DM_impurity}.
Here, we focus on the \DM\ interaction~(DMI) \cite{Dzyaloshinskii, Moriya, DMsimple}. Its value has been experimentally estimated to $D\simeq0.08J$~\cite{DM_Bert}. Theoretical studies \cite{Elhajal, DM_Cepas, Moi_DM, DM_Fritz, DM_cluster, SFMFT_DM, Hering17_PRB.95.054418} have concluded that a transition occurs between a small-$D$ disordered phase and a $\mathbf q=0$  N\'eel state at $D \gtrsim 0.1J$.
But the precise nature of the disordered phase at small $D$ is still unclear.

Here we describe a new chiral SL within the framework of Schwinger-boson mean-field theory (SBMFT) as a strong candidate for the phase realized in herbertsmithite.
The state has a unique property: the bottom of the spin excitation continuum is flat over an extended quasi-circular region of the Brillouin zone.
We compute the dynamical structure factor and confront it with data of Han {\it et al.}~\cite{Nature_DM} and with the theory by Punk {\it et al.}~\cite{SBMFT_Nature}.

\textit{\DM\ interaction.}
The DMI \cite{Dzyaloshinskii, Moriya} is a consequence of spin orbit coupling and comes from a broken mirror symmetry.
It is characterized by vectors $\mathbf D_{ij} = 2J\theta_{ij}\mathbf d_{ij}$ on oriented links ($\mathbf D_{ij}=-\mathbf D_{ji}$), where $\mathbf d_{ij}=\mathbf d_{ji}$ has unit length.
The total spin interaction on link $(ij)$ is \cite{DMsimple, suppMat}
\begin{equation}
\label{eq:Hlink}
h_{ij} = J \, \mathbf S'_i\cdot \mathbf S'_j\,,
\end{equation}
where $\mathbf S'_i$ and $\mathbf S'_j$ are obtained from the original spins by rotations around the $\mathbf d_{ij}$ axis with angles $\theta_{ij}$ and $-\theta_{ij}$, respectively.
In the following, we set $J=1$.
The Hamiltonian is the sum over nearest-neighbor link energies,
\begin{equation}
  \label{eq:H}
  H = \sum_{\langle i,j\rangle} h_{ij}\,.
\end{equation}
When the composition of these rotations around a lattice loop is identity, then all nontrivial angles $\theta_{ij}$ can be removed by a unitary transformation and the spectrum is unaffected \cite{Kaplan, Essafi}.
Otherwise, the effect of nonzero $\theta = |\theta_{ij}|$ depends on the geometry of the lattice.
For example, on the antiferromagnetic square lattice, spins are unfrustrated and $\theta$ increases the ground state energy by introducing frustration. On the kagome lattice, the presence of loops with an odd number of sites (triangles) maximally frustrates antiferromagnetic interactions. In this case, a nonzero $\theta$ decreases the ground state energy by reducing frustration.

\begin{figure}
 \includegraphics[width=0.3\textwidth]{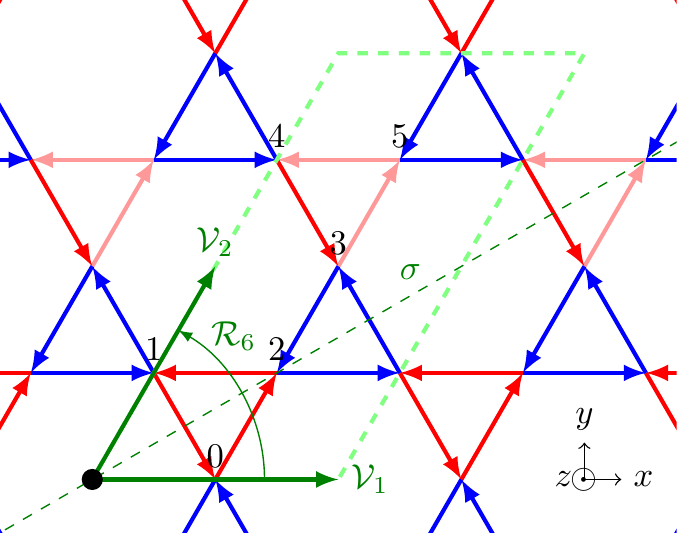}
 \caption{
 The kagome lattice and its symmetries (in dark green).
 The orientation of the $\mathbf D_{ij}$ (Dzyaloshinskii-Moriya) vectors on the directed links is out of plane.
 The unit cell of the \textit{Ansatz} (light green) contains 6 sites.
 Red and blue arrows represent first-neighbor links wearing mean-field parameters $\mathcal A_{ij}$ and $\mathcal B_{ij}$,
 equal to $|\mathcal A|$ and $|\mathcal B| e^{i\phi_{\mathcal B}}$ on red links, and
 $|\mathcal A|e^{i\phi_{\mathcal A}}$ and $|\mathcal B| e^{i(1-2p_R)\phi_{\mathcal B}}$ on blue links,
 with an additional phase $p_1\pi$ on light red bonds.
\label{fig:DM_orientation}
}
\end{figure}

Using crystal symmetry considerations, we can restrict the set of possible $\mathbf D_{ij}$.
In herbertsmithite, it has constant modulus and is perpendicular to the $(ij)$ link.
Electron spin resonance measurements evaluated $\mathbf D_{ij}$ to be mainly perpendicular to the kagome plane and of order $D = |\mathbf D_{ij}| \simeq 0.08$ ($\theta\simeq 0.04$) \cite{DM_Bert}.
The direction of $\mathbf D_{ij}$ on a reference link fixes all the other directions (Fig.~\ref{fig:DM_orientation}).
The tripartite nature of the lattice implies a $\pi/3$ periodicity in $\theta$ (up to a sublattice-dependent spin rotation).
Since $\theta_{ij}$ and $-\theta_{ij}$ are equivalent up to a mirror reflection, we can limit our study to $0\leq\theta\leq\pi/6$.
The Hamiltonian of Eq.~\eqref{eq:H} breaks some symmetries of the pure Heisenberg model:
$\sigma$ (lattice mirror symmetry) and
SU(2) spin rotations.
The preserved symmetries (Fig.~\ref{fig:DM_orientation}) are generated by:
$\mathcal V_1$ and $\mathcal V_2$ (lattice translations),
$\mathcal R_6$ (lattice rotation of order 6),
$\mathcal \sigma S_{\pi x}$ (mirror symmetry $\sigma$ combined with a spin rotation of $\pi$ around the $x$ axis), U(1) spin rotations around the $z$ axis, and $\mathcal T$ (time-reversal symmetry).

For classical spins, DMI immediately lifts the extensive ground state degeneracy of the Heisenberg model to the planar \qzero\ state of one of the two possible vector chiralities $\mathbf S_1 \land \mathbf S_2$ \cite{Elhajal} (but the scalar chirality $\chi_{123} = \mathbf S_1\cdot(\mathbf S_2\land \mathbf S_3)$ remains zero).
In the quantum $S=1/2$ model, a transition from a SL to this \qzero\ long-range order is expected at $D=D_c$ where $D_c\simeq 0.1$ \cite{DM_Cepas, Moi_DM, DM_Fritz}.
In the following, we elaborate on how to construct an elegant mean-field theory including DMI.

{\it SBMFT and chiral phases.}
In terms of the bosonic spinon $a_{i\alpha}$ of spin \mbox{$\alpha\in\{\uparrow,\downarrow\}$} on site~$i$, the spin operator reads as $\mathbf S_i = \frac12 a_{i\alpha}^\dag \bm \sigma_{\alpha\beta}a_{i\beta}$, where $\bm \sigma$ are the Pauli matrices.
The boson number is constrained to
\begin{equation}
  \label{eq:constraint}
  \sum_\alpha a_{i\alpha}^\dag a_{i\alpha} = 2S\,.
\end{equation}
In the mean-field theory, this constraint is enforced on average with the help of a Lagrange multiplier $\lambda$.

We define two operators on each link $(j,k)$:
\begin{eqnarray}
  A_{jk}&=&\frac{1}{2}\left( e^{-i\theta_{jk}}a_{j\uparrow}a_{k\downarrow}- e^{i\theta_{jk}}a_{j\downarrow} a_{k\uparrow} \right)\,,\\
  B_{jk}&=&\frac{1}{2}\left( e^{i\theta_{jk}}a_{j\uparrow}^\dag a_{k\uparrow}+e^{-i\theta_{jk}}a_{j\downarrow}^\dag a_{k\downarrow}\right)\,.
\end{eqnarray}
For $\theta=0$, $A_{jk}$ and $B_{jk}$ are invariant under global spin rotation.
For $\theta> 0$, this invariance is reduced to rotations around the $z$ axis. The link interaction, Eq.~\eqref{eq:Hlink}, can be written as
\begin{eqnarray}
\label{eq:hAB}
  h_{ij}
  &=&
  :B_{ij}^\dag B_{ij}:
  -A_{ij}^\dag A_{ij}
  \\
  \label{eq:hA}
  &=& S^2 - 2 A_{ij}^\dag A_{ij}\,,
\end{eqnarray}
where $:\,:$ means normal ordering.
Two different mean-field approximations can be developed using either the two parameters $\mathcal A_{ij} = \langle A_{ij}\rangle$ and $\mathcal B_{ij} = \langle B_{ij}\rangle$, and Eq.~\eqref{eq:hAB} ($\mathcal A \mathcal B$ formalism):
\begin{equation}
  h_{ij}^{\mathcal A\mathcal B} =
  \mathcal B_{ij}^* B_{ij}
  -\mathcal A_{ij}^* A_{ij}
  +H.c.
  - |\mathcal B_{ij}|^2 + |\mathcal A_{ij}|^2\,,
\end{equation}
or Eq.~\eqref{eq:hA} and the parameter $\mathcal A_{ij}$ only ($\mathcal A$ formalism).
Equations \eqref{eq:hAB} and \eqref{eq:hA} are identical in spin space when the constraint Eq.~\eqref{eq:constraint} is exactly imposed.
But in the enlarged Hilbert space of bosons where the constraint is only respected on average, they differ by a term $\propto (n_i-2S)(n_j-2S)$, related to the boson-number fluctuations.
The $\mathcal A$ formalism leads to inconsistencies, which have been discussed in detail for triangular and square lattices \cite{Symplectic_SBMFT, Trumper_AB_SBMFT}.
SBMFT has previously been used in attempts to describe DMI \cite{Moi_DM, DM_Fritz, DM_SBMFT}.
For the kagome lattice, however, this has only been done in the $\mathcal A$ formalism so far.

In order to reduce the total number of link parameters, we use the notion of projective symmetry group~\cite{Wen_PSG, PSG}.
This analysis has recently been extended to SLs where time reversal $\mathcal T$ can be broken, but where lattice symmetries (or their composition with $\mathcal T$) are preserved~\cite{PSG_chiral, kagome_SFMFT_chiral, PSG_chiral_fermions}.
Here, we restrict ourselves to \textit{Ans\"atze} respecting the symmetries of Eq.~\eqref{eq:H} in this sense (Fig.~\ref{fig:DM_orientation}).
We thus consider the generators
$\mathcal V_1$,
$\mathcal V_2$,
$\mathcal T^{p_\mathcal R}\mathcal R_6$, and
$\mathcal T^{p_\sigma} \sigma \mathcal S_{\pi x}$,
with $p_\sigma$, $p_\mathcal R=0$ or 1.
This results in 20 \textit{Ansatz} families listed in Table~\ref{tab:Ansaetze}.
In all these cases, $\mathcal A_{ij}$ and $\mathcal B_{ij}$ on a reference link are propagated to the entire lattice by rules that depend on $p_R$ and a parameter $p_1$ ($=0$ or~1) related to the presence of an additional $\pi$ flux through elementary tiles of the lattice.
For each family, an \textit{Ansatz} is characterized by two to four continuously adjustable parameters, corresponding to modulus and argument of $\mathcal A_{ij}$ and $\mathcal B_{ij}$ on the reference link, named $|\mathcal A|$, $\phi_{\mathcal A}$, $|\mathcal B|$, and $\phi_{\mathcal B}$.
These parameters are adjusted until self-consistent saddle point solutions are found.
In some families, $\phi_{\mathcal A}$ and $\phi_{\mathcal B}$ are restricted by discrete parameters $p_{\mathcal A}$ and/or $p_{\mathcal B}$ (=0 or 1).
The resulting link parameters are described in Fig.~\ref{fig:DM_orientation} and in the last two columns of Table~\ref{tab:Ansaetze}.

Note that the families shown in Table~\ref{tab:Ansaetze} possess common \textit{Ans\"atze}.
Clearly, $p_1$ discriminates two \textit{Ans\"atze} only when one of $|\mathcal A|$ or $|\mathcal B|$ is nonzero, while $p_R$ and $p_\sigma$ distinguish two \textit{Ans\"atze} with identical $p_1$ only when $\phi_\mathcal A$ or $\phi_\mathcal B$ is nontrivial ($\neq 0$ or $\pi$).
Some families can break $\mathcal T$ due to nontrivial $\phi_{\mathcal A}$ or $\phi_{\mathcal B}$.
In this case, fluxes through lattice loops take nontrivial values leading to nonzero scalar spin chiralities.
With Eq.~\eqref{eq:H}, we do not find any self-consistent solution with $\phi_{\mathcal B}\neq \pi$.
As a result, the $A_1, A_2$, and $A_3$ families never break $\mathcal T$.
Only a nontrivial $\phi_{\mathcal A}$ (allowed in the families $A_4$) may break it.

On the kagome lattice, scalar chirality $\chi_{123}$ is usually associated with elementary triangles.
In our framework, chiral \textit{Ans\"atze} with $p_R=0$ have uniform scalar chirality, while those with $p_R=1$ have chiralities of opposite sign on up and down triangles.
This implies that a nonzero global (i.e., a macroscopic) chirality is only possible for $p_R=0$. However, since $\chi_{123}$ is related to the imaginary part of $(|\mathcal B| e^{i\phi_\mathcal B})^3$, this is always trivial since we find $\phi_\mathcal B=\pi$.
Thus, none of our solutions exhibit a macroscopic chirality.

In the following, we shall call \textit{chiral state} any $\mathcal T$-breaking \textit{Ansatz}, even in the absence of a macroscopic chirality.
In such \textit{Ans\"atze}, some $\chi_{123}$ are nonzero, e.g.\ for three consecutive sites of a hexagon.
One could argue that the flux through a hexagon, $6\phi_\mathcal B(1-p_R)+p_1\pi$ (phase of $\mathcal B_{12}\mathcal B_{23}\dots\mathcal B_{61}$), is still trivial.
However, for loops with even parity, we can also consider the $\mathcal A$ flux $3\phi_\mathcal A+p_1\pi+\pi$ ($= \arg(\mathcal A_{12}(-\mathcal A^*_{23})\dots \mathcal A_{56}(-\mathcal A_{61}^*))$).
These two fluxes differ by their behaviour under $\mathcal R_6$ rotation:
the $\mathcal B$ flux is invariant, while the $\mathcal A$ flux changes sign.
Thus, a nontrivial $\mathcal B$ flux (only possible when $p_R=0$) characterizes a uniform chirality, $\chi_{123} = \chi_{234}$, while a nontrivial $\mathcal A$ flux (only possible when $p_R=1$) characterizes a staggered chirality, $\chi_{123}=-\chi_{234}$.
Note that, in the presence of a DMI, these fluxes contain $\theta$ in addition to the mean-field parameters, indicating a modified flux-chirality relation.

The existence of chiral phases as ground states \cite{Wen_Wilczek, PhysRevLett.70.2641, Hu2015_PRB.91.041124} is already evident in the classical limit: an infinitesimal antiferromagnetic third-neighbor interaction lifts the degeneracy of the kagome antiferromagnet to the nonplanar \cuboc\ state \cite{Regular_order}.
In the $\mathcal A \mathcal B$ formalism, this phase melts into a stable chiral $\mathbb Z_2$ SL (family $A_4(1,1)$ of Table~\ref{tab:Ansaetze}) at small spin~\cite{cuboc1}.
This example of spontaneous generation of scalar chirality is a strong motivation for taking chiral \textit{Ans\"atze} into account when solving the SBMFT problem with DMI.

\begin{table}
\renewcommand{\arraystretch}{1.1}
\begin{center}
 \begin{tabular}{|c||c|c|c|c|c|}
\hline
 & $p_R$ & $p_\sigma$ & $\phi_{\mathcal A}$ & $\phi_{\mathcal B}$\\
\hline
\hline
$A_1(p_1,p_{\mathcal A},p_{\mathcal B})$ &0 &0 &$p_{\mathcal A}\pi$ &$p_{\mathcal B}\pi$
\\
$A_2(p_1,p_{\mathcal A})$ &0 &1 &$p_{\mathcal A}\pi$ &n.t.
\\
$A_3(p_1,p_{\mathcal A})$ &1 &0 &$p_{\mathcal A}\pi$ &n.t.
\\
$A_4(p_1,p_{\mathcal B})$ &1 &1 &n.t.&$p_{\mathcal B}\pi$
\\
\hline
 \end{tabular}
\caption{\label{tab:Ansaetze}
Description of the 20 families of \textit{Ans\"atze} respecting all symmetries of kagome with DMI, up to time reversal.
$p_R$, $p_\sigma$, $p_1$, $p_{\mathcal A}$, and $p_{\mathcal B}$ (equal to 0 or 1) describe constraints on
the link parameters and their propagation to the entire lattice (Fig.~\ref{fig:DM_orientation}).
``n.t.'' means that the phase $\phi$ can take nontrivial values. The $A_1$ family has two adjustable parameters $|\mathcal A|$ and $|\mathcal B|$, whereas the others have three parameters ($\phi_{\mathcal A}$ or $\phi_{\mathcal B}$ in addition).
}
\end{center}
\end{table}


{\it Results.} We perform a numerical optimization of the parameters $|\mathcal A|$, $|\mathcal B|$, $\phi_{\mathcal A}$, and $\phi_{\mathcal B}$, using injection of the measured parameters until convergence, combined with a Brent algorithm to optimize the phases.
The mean-field energy is minimized with respect to $|\mathcal A|$ and $\phi_{\mathcal A}$, and maximized with respect to $|\mathcal B|$ and $\phi_{\mathcal B}$.
The Lagrange multiplier $\lambda$ is optimized each time a parameter is modified.

\begin{figure}
 \includegraphics[width=0.23\textwidth]{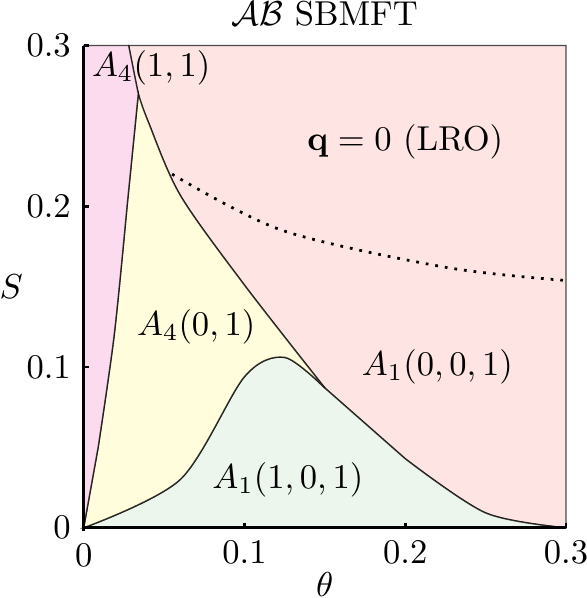}
 \includegraphics[width=0.23\textwidth]{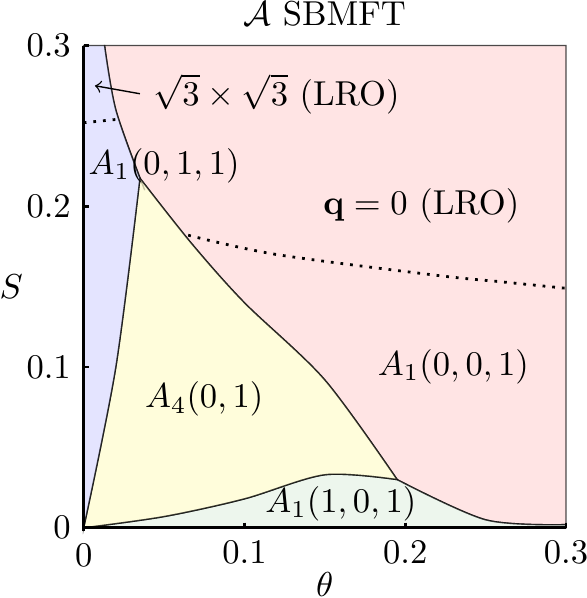}
 \caption{Phase diagram.
 The \textit{Ansatz} families with lowest self-consistent mean-field energy are indicated.
 LRO (above the dotted lines) means long-range order.
 The other phases are gapped $\mathbb Z_2$ SLs.
 }
\label{fig:phase_diagram}
\end{figure}

In SBMFT, the value of spin $S$ is a continuous parameter, given by the average number of bosons per site (Eq.~\eqref{eq:constraint}).
We optimize each \textit{Ansatz} family in Table~\ref{tab:Ansaetze}, and we select the one with lowest energy for fixed $S$ and $\theta$.
So constructed phases either exhibit N\'eel order or are gapped (chiral) $\mathbb Z_2$ spin liquids \cite{suppMat}.
Our results are summarized in Fig.~\ref{fig:phase_diagram} and discussed below.
For completeness, we also reproduce the phase diagram of Ref.~\cite{Moi_DM} in the $\mathcal A$ formalism, but here we include time-reversal breaking states as well (Fig.~\ref{fig:phase_diagram}(b)).

Let us discuss four special cases: $S\to\infty$, small $S$, $\theta=0$, and $\theta=\pi/6$.

(a) $S\to\infty$:
In the classical limit, we expect the mean-field solution to exhibit magnetic order through Bose-Einstein condensation of spinons.
For $\theta=0$, there is an extensive degeneracy, but the only N\'eel states that are reachable with our symmetric \textit{Ans\"atze} are the \textit{regular} ones, constructed in~\cite{Regular_order}.
Three of them belong to the ground state manifold: \qzero, \sqrtsqrt, and \cuboc.
They are obtained, respectively, from $A_1(0,0,1)$, $A_1(0,1,1)$, and $A_4(1,1)$ of Table~\ref{tab:Ansaetze}. All three \textit{Ans\"atze} approach the same energy, and they show the classical values of the mean-field parameters~\cite{PSG_chiral}. A nonzero DMI favors $A_1(0,0,1)$ (i.e., \qzero) consistent with a classical analysis.

(b) small $S$:
In the $\mathcal A$ formalism and following \mbox{Tchernyshyov} {\it et al.} \cite{flux}, this limit can be solved through an expansion in $S$. In the presence of a DM flux, defined as the usual flux $\arg(\mathcal A_{ij}(-\mathcal A^*_{jk})\dots \mathcal A_{lm}(-\mathcal A_{mi}^*))$ plus $\theta_{ij}+\theta_{jk}+\dots+\theta_{mi}$, we find that the expansion of the energy to order 8 agrees with the right panel of Fig.~\ref{fig:phase_diagram}, up to $S\simeq0.15$.

(c) $\theta=0$ (pure Heisenberg case \cite{pureHeisenberg}):
As shown previously \cite{cuboc1}, we find $A_4(1,1)$ (i.e., \cuboc) in the $\mathcal A\mathcal B$- and $A_1(0,1,1)$ (i.e., $\sqrt3\times\sqrt3$) in the $\mathcal A$ formalism as the lowest-energy phase.

(d) $\theta=\pi/6$:
Classically, the \qzero\ N\'eel state with well chosen vector chirality minimizes the link energy and is the unique ground state.
The Hamiltonian Eq.~\eqref{eq:H} is thus unfrustrated.
It is equivalent to the $XXZ$ model with ferromagnetic $XX$ coupling.
In this model, quantum Monte Carlo simulations found a superfluid phase~\cite{Kag_XXZ_ferro}.
As a consequence of the absence of frustration, $|\mathcal B|=0$, and the two formalisms are equivalent (similar to the square lattice).
$A_1(0,0,1)$ is thus the lowest-energy state for any value of spin (Fig.~\ref{fig:phase_diagram}).

Five of the twenty \textit{Ansatz} families of Table~\ref{tab:Ansaetze} appear as ground states of our model in the range of parameters of Fig.~\ref{fig:phase_diagram}.
Two of them break $\mathcal T$ and were absent in $\mathcal T$-symmetric investigations~\cite{Moi_DM}.
In addition to the chiral \textit{Ansatz} $A_4(1,1)$ already discussed for $\theta=0$~\cite{cuboc1}, a new chiral phase is found here, both in the $\mathcal A\mathcal B$- and in the $\mathcal A$ formalism: the $A_4(0, 1)$ phase.

Since SBMFT contains unphysical boson number fluctuations, some care must be taken in the interpretation of these results \cite{Trumper_AB_SBMFT}.
However, we consistently obtain the new phase in two formalisms ($\mathcal A$ and $\mathcal A\mathcal B$), where the fluctuations are treated differently.
This is an indication that the phase is robust and that it can survive an enforcement of the strict constraint Eq.~\eqref{eq:constraint}.

\begin{figure}
 \includegraphics[height=0.19\textwidth]{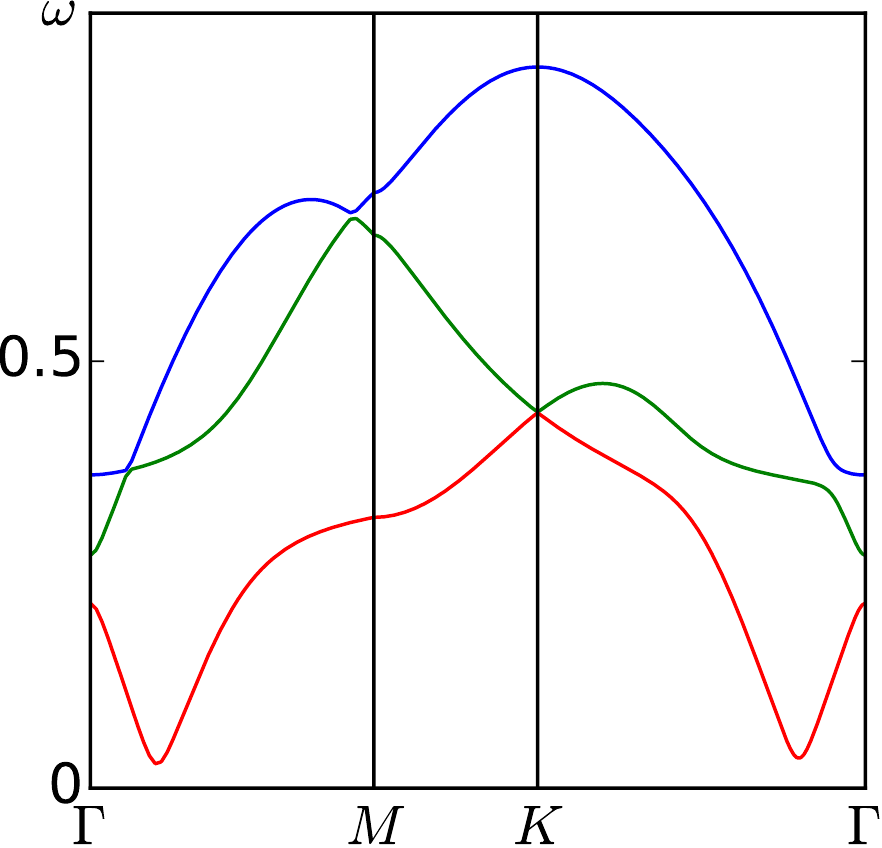}
 \includegraphics[height=0.2\textwidth]{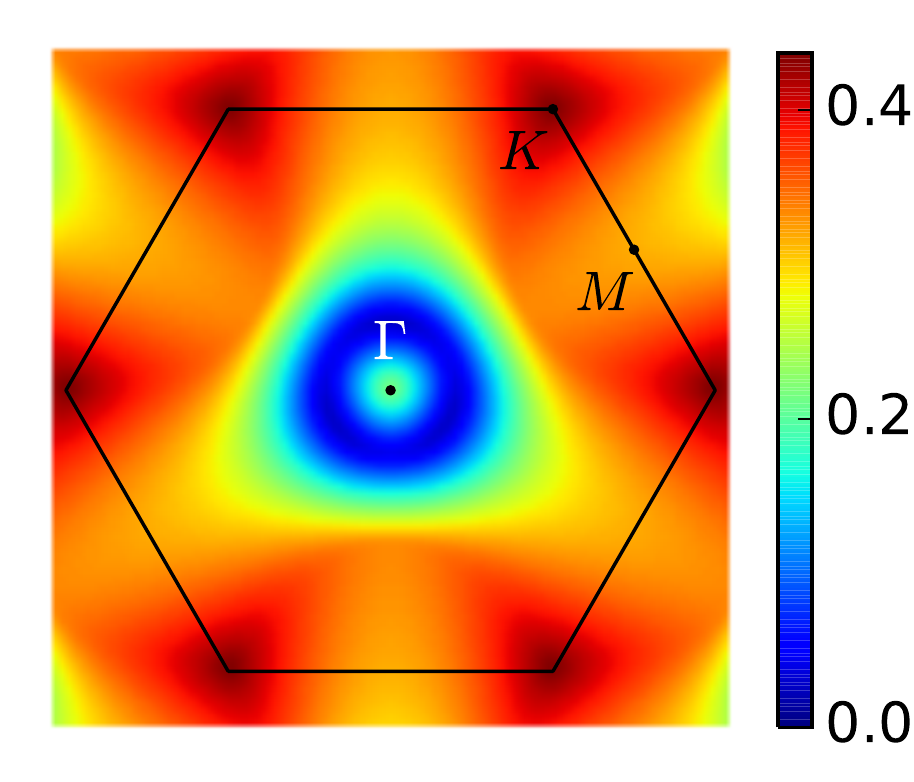}
 \caption{Typical spinon spectra in the $A_4(0,1)$ phase for small DMI and small spin (here $\theta=0.01$ and $S=0.5$; $\mathcal A\mathcal B$ formalism).
 The left panel shows the spinon energies along a cut, the right one shows the lowest band in the full Brillouin zone 
 (with the characteristic ring of low-energy excitations).
 \label{fig:spinons}
 }
\end{figure}

The new chiral phase $A_4(0,1)$ is separated from adjacent phases by first order phase transitions.
Because of the hysteresis phenomenon, its domain of metastability is larger than shown in Fig.~\ref{fig:phase_diagram} \cite{suppMat}.
It is notably metastable for $\theta=0$ up to $S\simeq0.65$ in the $\mathcal A\mathcal B$ formalism, and up to $S\simeq0.3$ in the $\mathcal A$ formalism.
In its entire domain of metastability, this phase has a closed curve of minimal-energy spinons in the Brillouin zone (Fig.~\ref{fig:spinons}).
To our knowledge, this intriguing property is unprecedented: previously studied gapped phases have sharply localized minima in the spinon spectrum~\cite{suppMat}.

\begin{figure}
 \includegraphics[width=0.48\textwidth]{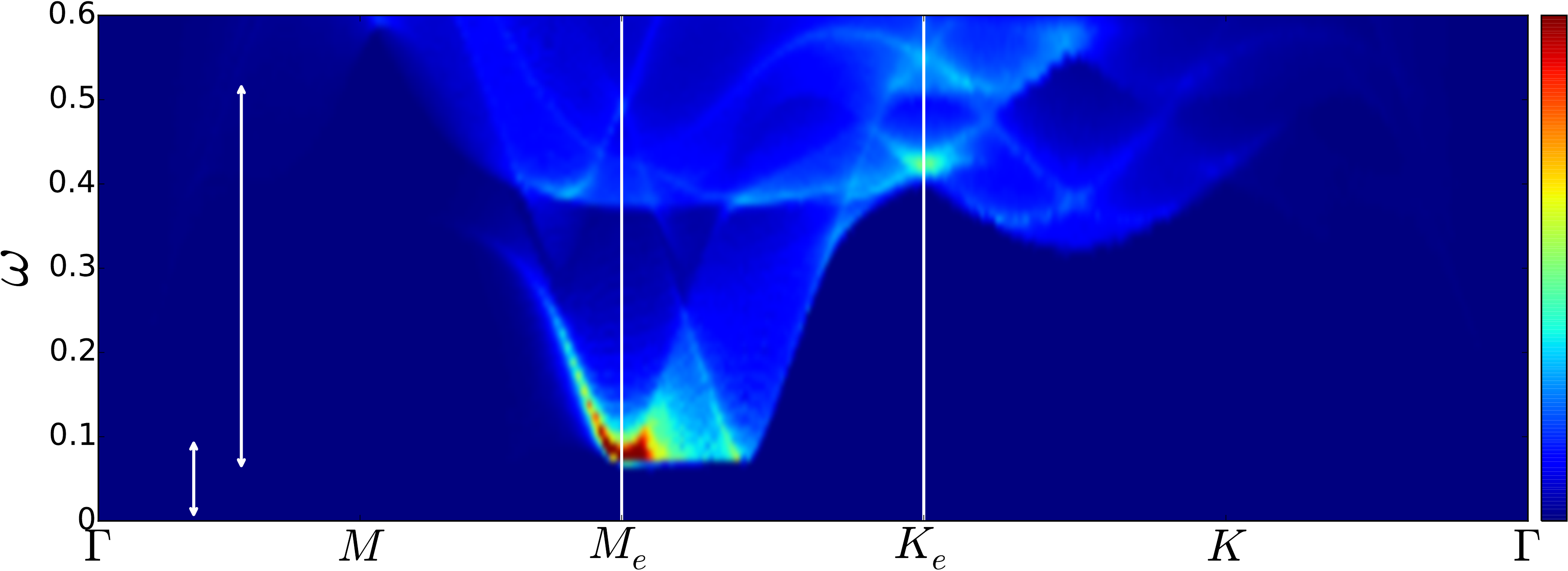}
 \includegraphics[width=0.23\textwidth]{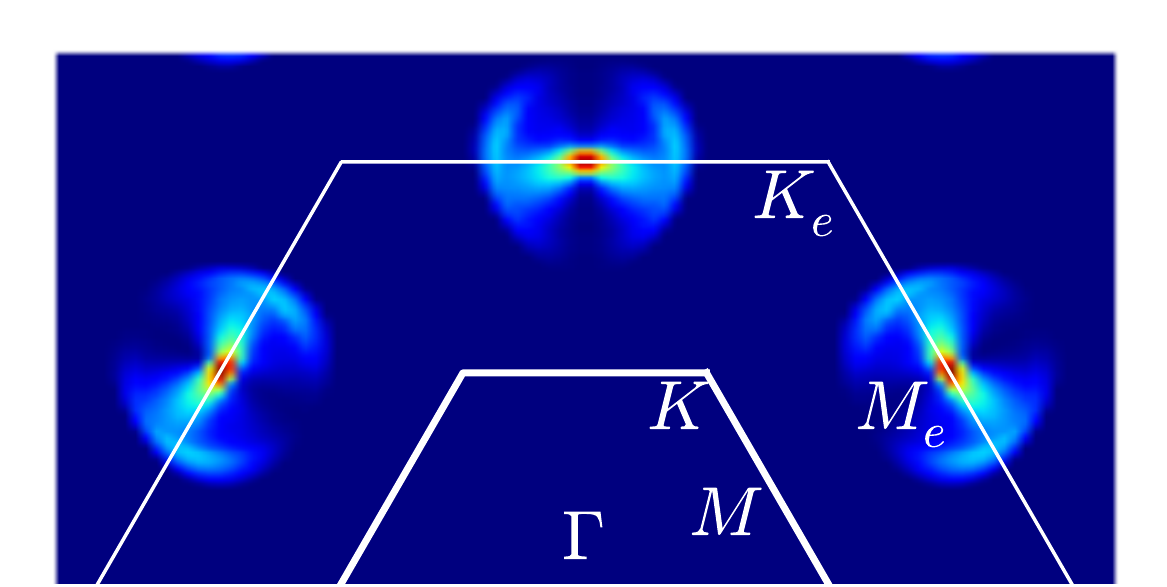}
 \includegraphics[width=0.23\textwidth]{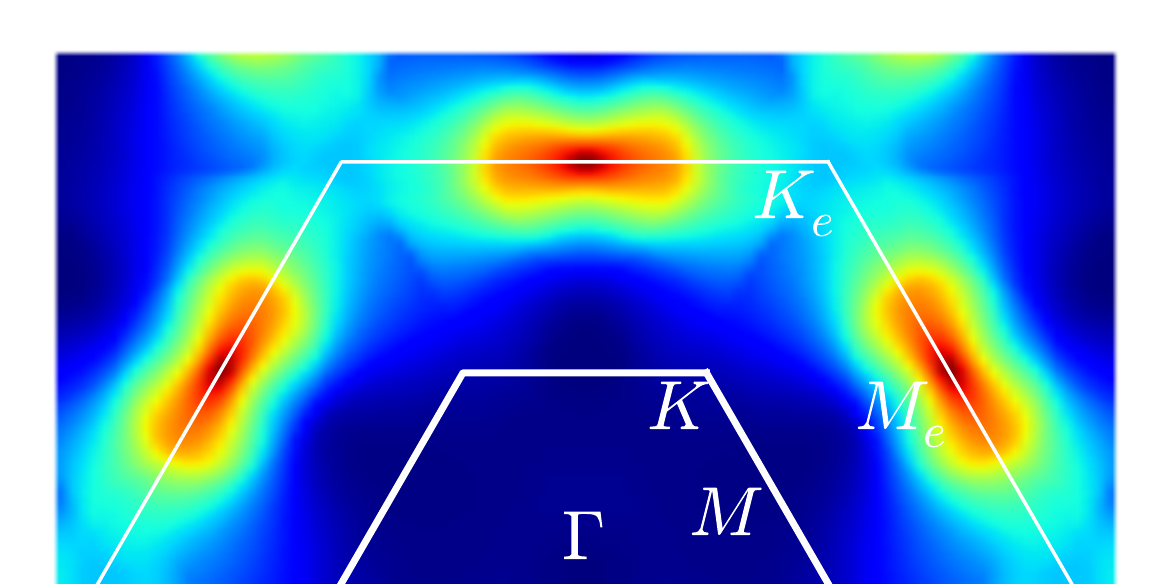}
 \caption{Top: dynamical structure factor $S(\mathbf q, \omega)$ of the $A_4(0,1)$ phase 
 (same parameters as in Fig.~\ref{fig:spinons}).
 The spin gap is $0.06 J$.
 Bottom left: $S(\mathbf q, \omega)$ integrated up to $\omega = 0.1J$.
 Bottom right: integrated over $0.06J<\omega<0.52J$.
 See \cite{suppMat} for similar results using different model parameters.
 }
\label{fig:dyn_struct_fact}
\end{figure}

Inelastic neutron scattering measures the dynamical structure factor $S(\mathbf q, \omega)$, i.e., the Fourier transformed space-time spin-spin correlations.
In SBMFT, $S(\mathbf q,\omega$) is nonzero when two spinons have the sum of their wave vectors equal to $\mathbf q$ and of their energies equal to $\omega$.
In previously studied SLs, the low-lying spin excitations consist of combinations of a spinon at a singular spectral minimum with one in the low-energy branch.
This leads to a high-intensity spot at the bottom of the excitation continuum, located at the Bragg peak of the corresponding classical phase, and to a strong dispersion away from this spot \cite{HalimehPunk16_PRB.94.104413}.
In contrast, for the new phase proposed in this article, any combination of two spinons on the minimum-energy curve has the same energy equal to twice the spinon gap.
This leads to a spin excitation spectrum that is flat in an extended region of the Brillouin zone (Fig.~\ref{fig:dyn_struct_fact}).

Inelastic neutron data on single-crystal herbertsmithite revealed a surprising spreading of intensity over a wide range of wave vectors at very low energy (0.75~meV $\simeq 0.04J$) \cite{Nature_DM}.
The low-energy structure factor of the $A_4(0,1)$ phase, Fig.~\ref{fig:dyn_struct_fact}, indeed shows analogies with these results in the correct energy range, but with stronger intensity variations.
The two bottom panels of Fig.~\ref{fig:dyn_struct_fact} can be compared with Figs.~(1c) and (1d) of \cite{Nature_DM}, respectively.

An attempt to explain Han's results by including vison excitations in the $A_1(0,0,1)$ phase was realized by Punk {\it et al.}~\cite{SBMFT_Nature}.
It was shown that this can indeed spread out the signal.
But the energy scale of $A_1(0,0,1)$ was not naturally consistent with experiment (theoretical results at $\omega=0.37J$ were compared to an experimental cut at $\omega=0.044J$).
In the new $A_4(0,1)$ phase, the energy scales are consistent, and we expect that adding visons can give a fairly convincing agreement with experiment.

{\it Conclusion.} We have realized a SBMFT study of the kagome antiferromagnet with DMI, including time-reversal symmetry breaking \textit{Ans\"atze}.
One of the self-consistent solutions has particularly interesting features: it is a small-gap $\mathbb Z_2$ SL with a finite density of minimal-energy excitations, stable in an extended region of the phase diagram (Fig.~\ref{fig:phase_diagram}).
Its dynamical structure factor fairly well reproduces the inelastic neutron scattering measurements on herbertsmithite~\cite{Nature_DM}: intensities around $\omega = 0.04J$ are obtained over a region of the Brillouin zone that is larger than in previously proposed $\mathbb Z_2$ SLs (Fig.~\ref{fig:dyn_struct_fact}).
Inclusion of visons \cite{SBMFT_Nature} in the model will be a promising step towards a faithful correspondence between theory and experiment.

{\it Acknowledgements.} L. M. would like to thank K. Penc for interesting discussions on the DMI.
This work was supported in part by grant ANR-12-BS04-0021 (France).

\bibliography{article_DM_chiral}

\end{document}